\documentclass[a4paper,10pt]{article}
\usepackage[english]{babel} % per lingua. TeXnic dà warning per misteriosi motivi...
\usepackage{graphicx,color}
\usepackage{latexsym}
\usepackage{amsmath,amsfonts,amssymb, amsthm}
\usepackage{physics}     % convenient physics shortcuts
\usepackage{braket}
\usepackage{varwidth}
\usepackage{hyperref}

\title{Tunnelling processes for Hadamard states through a 2+1 dimensional black hole and Hawking radiation}

\author{Francesco Bussola$^{1,2,a}$, Claudio Dappiaggi$^{1,2,b}$
	\vspace{4mm}\\
	{\small $^1$ Dipartimento di Fisica, Universit{\`a} di Pavia -- Via Bassi 6, 27100 Pavia, Italy.}\vspace{2mm}\\
	{\small $^2$ INFN, Sezione di Pavia -- Via Bassi 6, 27100 Pavia, Italy.}\vspace{2mm}\\
	{\footnotesize  ~$^a$ francesco.bussola@pv.infn.it~,~$^b$ claudio.dappiaggi@unipv.it~}}

\begin{document}
\maketitle

\begin{abstract}
We analyse the local behaviour of the two-point correlation function of a quantum state for a scalar field in a neighbourhood of a Killing horizon in a $2+1$-dimensional spacetime, extending the work of Moretti and Pinamonti in a $3+1$-dimensional scenario. In particular we show that, if the state is of Hadamard form in such neighbourhood, similarly to the $3+1$-dimensional case, under a suitable scaling limit towards the horizon, the two-point correlation function exhibits a thermal behaviour at the Hawking temperature. Since the whole analysis rests on the assumption that a Hadamard state exists in a neighbourhood of the Killing horizon, we show that this is not an empty condition by verifying it for a massive, real scalar field subject to Robin boundary conditions in the prototypic example of a three dimensional black hole background: the non-extremal, rotating BTZ spacetime. 
\end{abstract}

% INTRODUCTION

\section{Introduction}

Quantum field theory on curved backgrounds aims at exploring the behaviour of quantum systems in the presence of a fixed,
classical metric. In this setting, particular attention has been devoted to the case when the geometry of the spacetime describes a black hole. Aside from the intrinsic importance of these gravitating systems, they played a paramount role in revealing new unexpected, quantum phenomena, such as the emission of Hawking radiation from a black hole \cite{Hawking:1974sw}.

Since its first computation, many alternative derivations of such phenomenon have been proposed, though eventually, a good deal of attention has been devoted to studying the interplays between radiating black holes and the local properties of the underlying quantum field in a neighbourhood of a point on the event horizon. Such viewpoint culminated in the seminal paper by Parikh and Wilczek \cite{PW:2000} in which they associate the arise of thermal effects to a tunnelling process through the horizon, see also \cite{Kraus:1996,Srinivasan:1999,SPS:2002,P:2002}.

We will not dwell into the details of these papers, rather we will focus on the viewpoint, first advocated and thoroughly analysed in \cite{Moretti:2010qd}. Herein, the main message has been to show that the arise of a thermal behaviour from a tunnelling process, as proposed in \cite{PW:2000}, can be ascribed only to the leading singular structure of the two-point correlation function of a Hadamard state for the underlying field theory, which is taken to be for simplicity a real, massive scalar field.
 
The Hadamard condition has been fist introduced as a selection criterion for physically admissible quantum states, since on the one hand it guarantee that the short-distance behaviour of the state coincides with that of the Poincar\'e vacuum, while, on the other hand, it ensures that the quantum fluctuations of all observables are finite and that there exists a covariant scheme to construct Wick polynomials, \cite{Radzikowski:1996pa,Radzikowski:1996ei,Khavkine:2014mta, Kay:1988mu}.

Subsequently, in a very influential paper by Haag and Frendehagen \cite{Fredenhagen:1989kr}, it has been shown that the Hadamard condition is deeply connected to the presence of Hawking radiation, although, in this paper, the focus was on the role of radiative modes.
Only in \cite{Moretti:2010qd} the attention has been switched towards the analysis of the interplay between Hadamard states and the thermal properties of field theoretical models in a neighbourhood of a Killing horizon. The key ingredient in this paper has been the integral kernel of a two-point function $\omega_2(x,y)$.
In particular, using the structural properties of a Killing horizon \cite{Kay:1988mu,Racz:1992bp} and provided that $\omega_2$ is of Hadamard form ina  neighborhood of the horizon, it has been proven that, independently of the choice of such bi-distribution, $\omega_2$ acquires a thermal spectrum with respect to the notion of time and energy associated with the Killing field, generating the horizon. This result is obtained by smearing $\omega_2$ with two test functions whose support is close to the horizon. 
More precisely, if the points $x$ and $y$ of the integral kernel of $\omega_2$ lie on the same side of the horizon, the Fourier transform of the two-point function displays a Bose factor at the Hawking temperature. Conversely, if $x$ and $y$ are kept at the opposite
sides of the horizon, the resulting spectrum agrees with the transition probability between two weakly coupled reservoirs which are in thermal equilibrium at the Hawking temperature, in perfect agreement with the predictions of \cite{PW:2000}.

Yet a close scrutiny of \cite{PW:2000} unveils that their analysis can be applied to any spacetime with a Killing horizon, regardless of the dimension of the background, while the results of \cite{Moretti:2010qd} are based on the leading singular behaviour of the local form of the two-point function of a Hadamard state in four spacetime dimensions. This prompts the question whether the above conclusions can be extended to any dimension and in this paper we investigate the case of a three-dimensional manifolds with a Killing horizon, showing that the results of \cite{Moretti:2010qd} are still valid regardless of the different singular structure of the Hadamard states. The interest towards this scenario is motivated also by the fact that the analysis of \cite{Moretti:2010qd} rests on the assumption that there exists a state which is of Hadamard form in a neighbourhood of the Killing horizon. While this is not an obstruction when the underlying manifold is globally hyperbolic, since the existence of Hadamard sates is guaranteed a priori thanks to a deformation argument \cite{Fulling:1981cf}, the situation is completely different when the background does not enjoy such property. This is especially relevant in $2+1$ dimensions, when considering the prototypic example of a $2+1$-dimensional black hole spacetime with a Killing horizon: the BTZ solution of the Einstein's equations with a negative cosmological constant \cite{Banados:1992wn,Banados:1992gq}. This is a stationary, axisymmetric manifold which is locally isometric to AdS$_3$, the three-dimensional anti-de Sitter spacetime, and it is not globally hyperbolic. At the level of field theory this has far reaching consequences since there is no guarantee that one can identify a state for a free, massive, real scalar field theory on BTZ which is of Hadamard form in a neighborhood of the outer horizon. This entails that the generalization of the work of \cite{Moretti:2010qd} might not be testable in the basic example of a three dimensional black-hole, albeit the tunnelling processes and Hawking radiation on BTZ have been investigated from different, complementary viewpoints by several research groups, see {\it e.g.} \cite{Medved:2001ca,Liu:2005hj,Angheben:2005rm,Wu:2006pz}. Hence, as a last step in this work, we argue that this is not the case since one can exploit that BTZ can be constructed directly from AdS$_3$ via a suitable periodic identification of points in order to individuate, starting from the ground state for a massive, real scalar field on the universal cover of AdS$_3$ with boundary conditions of Robin type \cite{Dappiaggi:2018xvw}, a Hadamard state in the whole BTZ spacetime.

\vskip .3cm

\noindent This paper is structured as follows. Section \ref{sec:tunnel} offers a brief review of the method proposed by \cite{Moretti:2010qd}
and sets some notations and conventions. In addition we generalize the results of \cite{Moretti:2010qd} to a 
$2+1$-dimensional spacetime equipped with a Killing horizon.\\
In Section \ref{sec:BTZ} we test our general analysis in the distinguished case
of a scalar field on BTZ spacetime. Subsection \ref{subsec:geom} lists some notable geometrical properties of such background, in particular that it possesses two bifurcate horizons, an inner and an outer one. Finally in Subsection \ref{subsec:KMS} we consider a massive real scalar field theory on BTZ spacetime showing that, for boundary conditions of Robin type, it admits a Hadamard state in a neighbourhood of the outer horizon, which is the key prerequisite to apply the reasoning of Section \ref{sec:tunnel}. This statement combines two distinct results. The first, based on \cite{Bussola:2017}, is the explicit construction, via a mode decomposition, of the ground state for a massive real scalar field with Robin boundary conditions, living in the exterior region of the outer horizon of a BTZ spacetime. This allows in turn to construct explicitly states obeying the KMS condition at arbitrary temperature. The second result is based on \cite{Dappiaggi:2018xvw} and it refers to the proof of the existence of a one parameter family of maximally symmetric states, locally of Hadamard form, for a real, massive scalar field on the universal cover of anti-de Sitter spacetime. Each of these states is characterized by the choice of a boundary condition of Robin type. 

% HR AS TUNNELLING

\section{Tunnelling processes in $2+1$ dimensions}
\label{sec:tunnel}
In this section we review the tunnelling process, first devised by Parikh and Wilczeck in \cite{PW:2000}, here presented with the local approach as proposed by Moretti and Pinamonti in \cite{Moretti:2010qd}. As we will show, although developed in connection to four dimensional spacetimes, this viewpoint is well suited to be generalized to arbitrary spacetime dimensions, $2+1$ in particular.

\subsection{General setting}\label{sec:setting}

In the following, let $(\mathcal{M},g)$ be a smooth, three dimensional, connected Lorentzian manifold. In addition, following \cite[Def. 2.1]{Moretti:2010qd} we assume that there exists an open subset $\mathcal{O}\subset \mathcal{M}$ such that
\begin{itemize}
	\item[(a)] there exists $K\in\Gamma(T\mathcal{O})$ for which $\mathcal{L}_Kg=0$,
	\item[(b)] the orbits of $K$ lying in $\mathcal{O}$ are diffeomorphic to an open interval of $\mathbb{R}$,
	\item[(c)] there exists a two dimensional, connected submanifold $\mathcal{H}\subset\mathcal{O}$, dubbed {\em local Killing horizon}, invariant under the action of the group of local isometries generated by $K$,
	\item[(d)] $K|_{\mathcal{H}}$ is lightlike and the intersection between $\mathcal{H}$ and the integral curves of $K$ identifies a smooth 2-dimensional submanifold of $\mathcal{H}$,
	\item[(e)] $\kappa$, the surface gravity of $\mathcal{H}$ is a non vanishing, positive constant.
\end{itemize}

Observe that the hypotheses above encompass the case of $(\mathcal{M},g)$ being endowed with a Killing field $K$ generating a bifurcate Killing horizon, see also Section \ref{sec:BTZ}. In this case $K$ is vanishing on a two-dimensional acausal submanifold $\mathcal{B}$ and it is lightlike on two $K$-invariant null submanifolds, $\mathcal{H}_{+}$ and $\mathcal{H}_{-}$ such that $\mathcal{B}=\mathcal{H}_{+}\cap \mathcal{H}_{-}$. Most notably  any neighbourhood of a point $p\in\mathcal{H}_{\pm}$, not intersecting $\mathcal{B}$, satisfies the geometric hypotheses. From now we shall assume to work in this setting. This is not at all a restriction since, whenever we consider $\mathcal{O}\subset\mathcal{M}$ fulfilling the hypotheses above, it is possible to deform smoothly $(\mathcal{M},g)$ so for a bifurcate Killing horizon to exist. On account of our interest in evaluating quantities only defined in $\mathcal{O}$, this procedure is harmless. 
 
In view of \cite{Kay:1988mu,Racz:1992bp} we can consider in $\mathcal{O}$ a distinguished coordinate patch $(V,U,x^3)$. Here $U$ denotes an affine parameter along the null geodesics, which are the integral lines of $K$ with the origin fixed to lie at $\mathcal{B}$. At the same time $V$ is the affine parameter, with origin at $\mathcal{B}$, of the integral curves of the future-pointing lightlike vector field $n_{\mathcal{H}_+}$ of $\mathcal{H}_+$. This is built as the parallel transport of $n$, the unique, future pointing, lightlike vector at $\mathcal{B}$ such that with $g(n,-\pdv{U})=-\frac{1}{2}$. The remaining unknown, $x_3$ denotes any, but fixed coordinate defined on an open neighbourhood of a point lying in $\mathcal{B}$. Every such chart defines per restriction a counterpart on $\mathcal{O}$.

With respect to the coordinates $(U,V,x_3)$, it turns out that the Killing vector field $K$ in $\mathcal{O}$ reads
\begin{equation}\label{eq:K}
K=K^1\pdv{V} + K^2 \pdv{U} + K^3\pdv{x^3}
\end{equation}
and, if there exists $\mathcal{O}^\prime\subset\mathcal{O}$ such that its closure is compact and such that $p\in\mathcal{O}^\prime$, then $K^1(p)=-\kappa V + V^2 R_1(p)$, $K^2(p)=\kappa U + V^2 R_2(p)$, $K^3(p)=V R_3(p)$. Here $R_1$, $R_2$, $R_3$ are bounded smooth functions on $\mathcal{O}^\prime$, {\it cf.} \cite[Prop. 2.1]{Moretti:2010qd}, while $\kappa$ is the surface gravity. In addition the line element of the metric $g$ reads
\begin{equation}\label{eq:horizong}
g_{\upharpoonright\mathcal{H}^{+}} = -\frac{1}{2}dU\otimes dV - \frac{1}{2}dV\otimes dU + h(x_3) dx^3\otimes dx^3
\end{equation}
where $_{\upharpoonright\mathcal{H}^{+}}$ indicates that the metric is defined in a neighbourhood of $\mathcal{H}^+$, while $h$ is a strictly positive function depending only on $x_3$. Combining together \eqref{eq:K} with \eqref{eq:horizong} it turns out that, at the leading order in $V$, $g_{\upharpoonright\mathcal{H}^{+}}(K,K)=\kappa^2 UV+O(V^2)$. Since $U$ is per construction of definite sign in $\mathcal{O}$, say positive, it turns out that, by shrinking if necessary $\mathcal{O}$, we can separate it as the union of three disjoint regions: $\mathcal{O}=\mathcal{O}_s\cup\mathcal{O}_0\cup\mathcal{O}_t$ where $\mathcal{O}_0=\mathcal{O}\cap\mathcal{H}_+$, while 
\begin{equation}\label{eq:regions}
\mathcal{O}_s\doteq\{p\in\mathcal{O}\;|\;V(p)<0\}\quad\textrm{and}\quad\mathcal{O}_t\doteq\{p\in\mathcal{O}\;|\;V(p)>0\},
\end{equation}
where the subscripts $s$ and $t$ indicate that the vector field $K$ is spacelike and timelike in $\mathcal{O}_s$ and $\mathcal{O}_t$ respectively. Heuristically and comparing with standard scenarios such as Schwarzschild spacetime, we can divide a neighbourhood of a point at the horizon in an interior ($s$) and an exterior ($t$) region.

Having established \eqref{eq:horizong}, it is possible to study the properties of the geodesic distance, which will play a key role in the next sections. Since the outcome is a slavish adaptation to a three dimensional scenario of \cite[Prop. 2.1]{Moretti:2010qd}, we will not dwell into the details of a systematic proof. More precisely, let $p\in\mathcal{O}$ be a point of coordinates $(U,V,x_3)$, so that $p\in\mathcal{H}^+$ if and only if $V=0$. Then it holds that

\begin{enumerate}
	\item Let $\widetilde{\mathcal{O}}\subset\mathcal{O}^\prime$ be any geodesically convex neighbourhood of $\mathcal{H}^+$ and let $p,q\in\mathcal{H}^+\cap\widetilde{\mathcal{O}}$.
	Then, the squared geodesic distance between these points is
	\begin{equation}\label{eq:sigma}
	\sigma(x(p),x(q))\equiv\ell(x_3(p),x_3(q))\doteq\left(\int\limits_{x_3(p)}^{x_3(q)}d\lambda\ f(\lambda)\right)^2,
	\end{equation}
	where $x(p)$ ({\em resp.} $x(q)$) indicates the representation of the point $p$ ({\em resp.} of the point $q$) in terms of the coordinates $(U,V,x_3)$. Furthermore $f^2=h$, $h$ being the function appearing in \eqref{eq:horizong}, while $x_3(p)$ and $x_3(q)$ are respectively the evaluation of $p$ and $q$ along the coordinate $x_3$.	
	\item Let $p\in\widetilde{\mathcal{O}}$ and, for any but fixed, admissible value of the coordinates $U^\prime, V^\prime$, let $S_{U^\prime,V^\prime}$ be the collection of points $q$ lying in the cross section of $\widetilde{\mathcal{O}}$ at constant $V^\prime$ and $U^\prime$. For $\delta>0$, let
	\begin{equation}\label{eq:G_delta}
	G_\delta(p,V^\prime, U^\prime)=\{q\in S_{U^\prime,V^\prime}\;|\; \ell(x_3(p),x_3(q))<\delta^2\},
	\end{equation}
	$\ell$ being as in \eqref{eq:sigma}. Then $\delta$ can be chosen in such a way that the smooth map $G_\delta(p,V^\prime, U^\prime)\ni q\mapsto \sigma(x(p),x(q))$ has minimum in a unique point $q(p,V',U')$. As a consequence of \eqref{eq:sigma} $x_3(q)=x_3(p)$ if $p\in\mathcal{H}^+\cap\widetilde{\mathcal{O}}$, whereas in general, there exists three bounded functions $F_i$, $i=1,2,3$, depending smoothly on $x(p),U^\prime, V^\prime$ such that	
	\begin{align}\label{eq:taylor}
	\sigma(x(p),x(q))=\ell(x_3(p),x_3(q))-(U-U^\prime)(V-V^\prime)+R(x(p),V,V^\prime,U^\prime)
	\end{align}
	where $R(x(p),U^\prime,V^\prime)=F_1V^2+F_2V'^2+F_3VV^\prime$.
\end{enumerate}

\subsection{Two-point Correlation Functions and their Scaling Behaviour}\label{Sec:2-pt_scaling}

On top of the spacetime $(\mathcal{M},g)$ we consider for definiteness a real, scalar field $\Phi:\mathcal{M}\to\mathbb{R}$ whose dynamics is ruled by the Klein-Gordon equation
\begin{equation}\label{eq:KG}
P\Phi=\left(\Box-m^2-\xi R\right)\Phi=0,
\end{equation}
where $\Box$ is the D'Alembert wave operator built out of $g$, $m^2\geq 0$, $R$ is the Ricci scalar while $\xi\in\mathbb{R}$. This choice is motivated mainly for its connection with Section \ref{sec:BTZ}, although the following analysis is independent from the choice of a detailed free field theory relying only on the specific assumptions on the singular structure of the two-point correlation function. The details of the quantization of \eqref{eq:KG} are an overkilled topic which has been thoroughly discussed in the literature. Therefore we will not dwell in the details, referring an interested reader for example to \cite{Benini:2013fia}. For the sake of clarity, we outline only the main points. To start with we assume that \eqref{eq:KG} can be solved in terms of an advanced and a retarded fundamental solutions $G^\pm:C^\infty_0(\mathcal{M})\to C^\infty(\mathcal{M})$ such that $P\circ G^\pm=G^\pm\circ P= id|_{C^\infty_0(\mathcal{M})}$ and such that $\textrm{supp}(G^\pm(f))\subseteq J^\mp(\textrm{supp}(f))$ for all $f\in C^\infty_0(\mathcal{M})$. Under such premise, we can associate to the scalar field $\Phi$ an algebra of observables $\mathcal{A}(\mathcal{M})$. This is constructed from the following basic structures:
\begin{enumerate}
	\item $\mathcal{T}(\mathcal{M})=\bigoplus_{n=0}^\infty (C^\infty_0(\mathcal{M}))^{\otimes n}$, where $C^\infty_0(\mathcal{M})^{\otimes 0}\equiv\mathbb{C}$ be the universal tensor algebra of test-functions, endowed with the $*$-operation induced from complex conjugation,
	\item $\mathcal{I}(\mathcal{M})$ be the $*$-ideal of $\mathcal{T}(\mathcal{M})$ generated by elements of the form $Pf$ (equation of motion) and $f\otimes f^\prime - f^\prime\otimes f- i G(f,f^\prime)\mathbb{I}$ (CCR) where $f,f^\prime\in C^\infty_0(\mathcal{M})$ while $\mathbb{I}$ is the identity of $\mathcal{T}(\mathcal{M})$ and $G\doteq G^--G^+$. 
\end{enumerate}

Hence we set $\mathcal{A}(\mathcal{M})=\frac{\mathcal{T}(\mathcal{M})}{\mathcal{I}(\mathcal{M})}$, which is also called {\em algebra of observables}. Expectation values of observables can be computed in terms of correlation functions $\omega_n$ which are $\mathbb{C}$-linear maps on $C^\infty_0(\mathcal{M})^{\otimes n}$ compatible both with the relations defining the ideal $\mathcal{I}(M)$ and with the positivity and normalization requirements needed to define a full-fledged state for the underlying quantum theory, see for example \cite{Khavkine:2014mta}.
If one is interested in the local properties of the system in a submanifold $\widetilde{\mathcal{O}}$, one might define a local algebra of observable as the restriction of $\mathcal{A}(\mathcal{M})$ on $\widetilde{\mathcal{O}}$.

In between the plethora of admissible correlation functions, we will be focusing on the special subclass of those of Hadamard form, see \cite{Kay:1988mu}. More precisely, adopting the notation and conventions of \cite{Moretti:2001qh}, we consider a two-point correlation function which descends from a distribution $\omega_2\in\mathcal{D}^\prime(\mathcal{M}\times\mathcal{M})$ such that the restriction of its integral kernel to $\widetilde{\mathcal{O}}\times\widetilde{\mathcal{O}}$ reads
\begin{equation}\label{eq:had}
\omega_{2,\epsilon}(x,x^\prime)=\frac{\Delta(x,x^\prime)^{1/2}}{4\pi\sqrt{\sigma_\epsilon(x,x^\prime)}}+\mathit{w}(x,x^\prime) \ ,
\end{equation}
where $x,x^\prime$ stand for the coordinates of two arbitrary points in $\widetilde{\mathcal{O}}$, $\Delta\in C^\infty(\widetilde{\mathcal{O}}\times\widetilde{\mathcal{O}})$ is the Van Vleck-Morette determinant, a quantity built only out of the metric, while $\mathit{w}(x,x^\prime)$ is a smooth function on $\widetilde{\mathcal{O}}\times\widetilde{\mathcal{O}}$.  The function $\sigma_\epsilon(x,x^\prime):= \sigma(x,x^\prime)+2i\epsilon(T(x)-T(x^\prime))+\epsilon^2$, where $\sigma(x,x^\prime)$ is the squared geodesic distance between $x$ and $x^\prime$, $\epsilon>0$ while $T$ is any, but fixed time function. Recall that $\widetilde{\mathcal{O}}$ is a convex geodesic neighbourhood with non vanishing intersection with $\mathcal{H}^+$. 

Having introduced all ingredients needed, we can now focus on extending to the $2+1$-dimensional scenario the analysis of \cite{Moretti:2010qd}. Let us consider thus two one-parameter families of test functions $f_\lambda,f^\prime_\lambda\in C^\infty(\widetilde{\mathcal{O}})$, $\lambda\in\mathbb{R}$ such that both obey to the constraints
\begin{equation}\label{eq:test_functions}
f_\lambda(V,U,x_3)=\frac{1}{\lambda}f\left(\frac{V}{\lambda},U,x_3\right)\quad\textrm{and}\quad f=\frac{\partial F}{\partial V},\;F\in C^\infty_0(\widetilde{\mathcal{O}}).
\end{equation}
 In the following, taking into account \eqref{eq:had}, we shall evaluate  
\begin{gather}\label{eq:Stpf}
\lim_{\lambda\to 0^{+}}\omega_{2}(f_\lambda,f^\prime_\lambda)\, =\notag\\
\lim_{\lambda\to 0^{+}}\lim_{\epsilon\to 0^{+}}\int_{\widetilde{\mathcal{O}}\times \widetilde{\mathcal{O}}} \left(\frac{\Delta(x,x^\prime)^{1/2}}{4\pi\sqrt{\sigma_\epsilon(x,x^\prime)}}+\mathit{w}(x,x^\prime)\right) f_\lambda(x)f^\prime_\lambda(x^\prime)d\mu_g(x) d\mu_g(x^\prime)\ ,
\end{gather}
where $d\mu_g=\sqrt{h(x_3)}dx_3dU dV$ is the metric induced volume form on $\widetilde{\mathcal{O}}$. We observe that only the contribution to \eqref{eq:Stpf} of the singular part of the two-point correlation function is relevant since, being $\mathit{w}$ smooth, the integral with respect to either $V$ or $V^\prime$ vanishes in view of the assumptions on the test functions. To this end we need to introduce an auxiliary cut-off function built as follows. Let $\delta>0$ and let $G_\delta(p,V^\prime,U^\prime)$ be as in \eqref{eq:G_delta}. We define $G_\delta(p,V^\prime,U^\prime)\ni p^\prime\mapsto\chi_\delta(x,x^\prime)\ge 0$ where $x$ indicates the coordinates of $p$, while $x^\prime=(V^\prime,U^\prime,x^\prime_3)$ those of $p^\prime$, to be any smooth and compactly supported function subject to the constraint
\begin{equation*}
\chi_\delta(x,x^\prime)=
1 , \text{ for }0 \le \sqrt{\ell(x_3(p),x_3(p^\prime))}\le \frac{\delta}{2}+\frac{1}{2}\sqrt{\ell(x_3(p),x_3(q))} 
\end{equation*} 
Here $q$ refers to the unique point $q(p,V^\prime,U^\prime)\in\widetilde{\mathcal{O}}$, minimizing the function $G_\delta(p,V^\prime,U^\prime)\ni p^\prime\mapsto \sigma(x,x^\prime)$. Hence, for any pair of points $p,p^\prime\in\widetilde{\mathcal{O}}$ of coordinates $x,x^\prime$ respectively, we rewrite the contribution to the two-point correlation function coming from the singular part of \eqref{eq:had} as 
\begin{subequations}
\begin{align}\label{split1}
\int_{\widetilde{\mathcal{O}}\times\widetilde{\mathcal{O}}}
d\mu_g(x) d\mu_g(x^\prime)\, \frac{\Delta(x,x^\prime)^{\frac{1}{2}}f_\lambda(x)f'_\lambda(x^\prime)}{4\pi\sqrt{\sigma_\epsilon(x,x^\prime)}}\chi_\delta(x,x^\prime)+\\ \label{split2}
\int_{\widetilde{\mathcal{O}}\times\widetilde{\mathcal{O}}}
d\mu_g(x) d\mu_g(x^\prime)\, \frac{\Delta(x,x^\prime)^{\frac{1}{2}}f_\lambda(x)f^\prime_\lambda(x^\prime)}{4\pi\sqrt{\sigma_\epsilon(x,x^\prime)}}(1-\chi_\delta(x,x^\prime))  
\end{align}
\end{subequations}
where $d\mu_g$ stands from the volume form induced by \eqref{eq:horizong}. In the following our strategy is to evaluate separately the limits of the above two terms as $\lambda$ and $\epsilon$ tend to $0^+$.

\vskip .2cm

\noindent{\it Evaluation of \eqref{split2}:} Following the same argument as in \cite{Moretti:2010qd}, by shrinking if necessary $\widetilde{\mathcal{O}}$, it turns out that the integrand is nowhere singular, being actually jointly smooth in all variables including if $\epsilon=0$. Hence we can apply Lesbegue dominated convergence theorem to exchange both limits with the integrals. The end point is an integrand, which is the derivative with respect to $V$ and $V^\prime$ of a compactly supported smooth function. Hence the overall integral vanishes.

\vskip .2cm

\noindent{\it Evaluation of \eqref{split1}:} In this case we need to deal with the singularity due to $\sqrt{\sigma_\epsilon(p,p')}$ as $\epsilon\to 0^+$. To this avail, for every $p\in\widetilde{\mathcal{O}}$, let $\rho:G_\delta(p,V^\prime,U^\prime)\to[0,\infty)$ be such that
	\begin{equation}\label{eq:rho}
	\sqrt{\sigma(x,x^\prime)} = \sqrt{\rho(x^\prime)^2+\sigma(x,x(q))},
	\, .
	\end{equation}
	where $G_\delta(p,V^\prime,U^\prime)$ is defined in \eqref{eq:G_delta} while $q$ is the point therein in which, for fixed $p$, the function $\sigma$ attains its unique minimum.
	Furthermore, from now on, in each $G_\delta(p,V^\prime,U^\prime)$ we switch from the coordinates $(U^\prime,V^\prime,x^\prime_3)$ to $(U^\prime,V^\prime,\rho)$. Using \eqref{eq:rho} and the Taylor expansion \eqref{eq:taylor}, \eqref{split1} reduces to
\begin{equation}\label{eq:aux1}
\int_{\widetilde{\mathcal{O}}\times\widetilde{\mathcal{O}}}
d\mu_g(x) d\mu_g(x^\prime)\,
\frac{\Delta(x,x^\prime)^{1/2}\chi_\delta(x,x^\prime)f_\lambda(x)f'_\lambda(x^\prime)}{4\pi\sqrt{\rho^2+\ell(x_3,x^\prime_3)-(U-U^\prime-i\epsilon)(V-V^\prime-i\epsilon)+R(x,V',U')}} .
\end{equation}

To write this integral in a more manageable form, we observe that the denominator of \eqref{eq:aux1} is the derivative with respect to $\rho$ of $$\Xi(\rho,V,V^\prime,R)=\ln(\left(\rho^2+\ell(x_3(p),x_3(q))-(U-U^\prime-i\epsilon)(V-V^\prime-i\epsilon)+R(x,V',U')\right)^{\frac{1}{2}}+\rho),$$
where we indicate only the explicit dependence of those variable which will play a role in the following discussion. Furthermore it is convenient to rescale $(V,V^\prime)$ to $(\lambda V, \lambda V^\prime)$ so that, calling $\Delta_\lambda, R_\lambda, d\mu_{g_\lambda}$ the quantities transformed accordingly and considering the hypothesis that $f=\partial_V F$, $f^\prime=\partial_{V^\prime}F^\prime$, we get
	\begin{equation*}
	\int_{\widetilde{\mathcal{O}}_\lambda\times\widetilde{\mathcal{O}}_\lambda}	d\mu_{g_\lambda}(x) d\mu_{g_\lambda}(x^\prime)\frac{\Delta_\lambda(x,x^\prime)^{\frac{1}{2}}}{4\pi}\chi_\delta(x,x^\prime)\,\partial_V F(x)\,\partial_{V^\prime} F^\prime(x^\prime)\,
	\partial_\rho\Xi\left(\rho,\lambda V,\lambda V^\prime,R_\lambda\right).
\end{equation*}

Since $\rho$ has a domain of definition of the form $[0,\rho_0)$ we can integrate by parts in this variable so to deal with the contribution from $\partial_\rho\Xi$. Of the ensuing boundary terms, that due to $\rho_0$ vanishes being $F^\prime$ compactly supported, while that due to $\rho=0$ yields, up to an innocuous rescaling of $\epsilon$ as $\lambda\epsilon$
	\begin{equation}\label{eq:aux2}
	-\int_{\widetilde{\mathcal{O}}_\lambda\times\widetilde{\mathcal{O}}_\lambda}	d\mu_{g_\lambda}(x) dU^\prime dV^\prime \left.\sqrt{|g|}\right|_{\rho=0} \frac{\Delta_\lambda(x,x^\prime)^{\frac{1}{2}}}{4\pi}\,\partial_V F(x)\,\partial_{V^\prime} F^\prime(x^\prime)|_{\rho=0}\,
	\left(\Xi\left(0,V,V^\prime,\frac{R_\lambda}{\lambda}\right)+\ln\lambda\right),
	\end{equation}
where we have written $d\mu_g(x^\prime)=\sqrt{|g|}d\rho\,dU^\prime dV^\prime$ and we have implicitly used that, when $\rho=0$, $\ell(x_3,x^\prime_3)=0$. Since we are interested in taking both the limit of \eqref{eq:aux2} as $\lambda$ and $\epsilon$ tend to $0^+$, we observe that the term proportion to $\ln\lambda$ does not contribute. Since both $F$ is smooth and compactly supported, one can integrate by parts in $V$. The boundary terms vanish while, due to the derivative of $\Delta_\lambda^{\frac{1}{2}}$, the remaining integral is proportional to $\lambda\ln\lambda$.

At the same time, in order to evaluate the remaining term in \eqref{eq:aux2}, first of all we notice that, in view of Eq. \eqref{eq:taylor}, being $R$ quadratic in $V$ and $V^\prime$, there exists $C\in\mathbb{R}$ such that, for $\lambda$ sufficiently close to $0^+$, $|\lambda^{-1}R_\lambda|<C\lambda$. In addition, by direct inspection of \eqref{eq:horizong} it turns out that, when either $x$ or $x^\prime$ tend to the horizon and $\lambda\to 0^+$, $\sqrt{|\det g_\lambda|}\to \frac{1}{2}$ while $d\mu_g(x)=\frac{dUdV}{2}d\mu(x_3)$ where $d\mu(x_3)=\sqrt{h(x_3)}dx_3$. As a last ingredient observe that, at a distributional level, letting $z=w+iy$, it holds $\lim\limits_{y\to 0^+}\ln(z)=\ln|w|+i\pi(1-\Theta(w))$, $\Theta$ being the Heaviside function. Gathering all these data together, the remaining contribution from \eqref{eq:aux2} reads as in \cite{Moretti:2010qd}

	\begin{equation}\label{eq:final}
	-
	\frac{1}{32\pi}\lim_{\epsilon \to 0^+}
\int_{\widetilde{\mathcal{O}}\times\widetilde{\mathcal{O}}}
	\partial_V F(x)\partial_{V^\prime} F^\prime(x^\prime)
	\ln(
	-(U-U^\prime)(V - V^\prime-i\epsilon)
	)
	dU^\prime dV^\prime dU dV d\mu(x_3)
	\, ,
	\end{equation}
where we have restored the limit as $\epsilon\to 0^+$ and where we used that $\Delta_0=1$ when $x_3=x^\prime_3$ and $V=V^\prime=0$.

As a consequence we managed to rearrange \eqref{eq:Stpf} in the form \ref{eq:final} which is a useful expression to investigate the energy spectrum of the two-point correlation function, as seen by an observer moving along the curves generated by $K$ with respect to the associated Killing time $\tau$.

\subsection{Thermal spectrum of the correlation functions}

Having established the form \eqref{eq:final} for the two-point correlation function of a real, massive scalar field in $2+1$ dimensions, we can repeat verbatim the analysis of \cite{Moretti:2010qd} aimed at computing the energy spectrum of \eqref{eq:final} as seen by an observer the moves along the integral curves of $K$. We summarize the procedure. It calls for considering two test functions which are squeezed on the Killing horizon.  From an operational point of view this consists of considering the leading behaviour of \eqref{eq:final} as $V$ is close to $0$. To this end, a direct inspection of \eqref{eq:K} unveils that, calling $\tau$ the affine parameter of the integral curves of $K$, it holds that, up to an irrelevant additive constant
\begin{equation}\label{V_to_tau}
V=-e^{-\kappa\tau}\;\textrm{for V}<0,\quad V=e^{-\kappa\tau}\;\textrm{for V}>0.
\end{equation}

\noindent Two cases should be considered when analysing $\omega_2(x,x^\prime)$: both points $x,x^\prime$ lie in the exterior region $\mathcal{O}_t$, {\it cf.} \eqref{eq:regions}, or one lies in $\mathcal{O}_t$ and the second in $\mathcal{O}_s$, the interior region.

\vskip .2cm

\noindent{\em Case $1)$} Let us consider two $1$-parameter families of test functions $f_\lambda$ and $f^\prime_\lambda$ obeying \eqref{eq:test_functions}, up to the replacement of $\widetilde{\mathcal{O}}$ with $\mathcal{O}_t$. Inserting this hypothesis in \eqref{eq:final}  we can first integrate by parts both in $V$ and in $V^\prime$. On account of the compactness of the support of both $F$ and $F^\prime$, no boundary term contributes to the result. Eventually, replacing $V$ with \eqref{V_to_tau}, we obtain
	\begin{equation*}
	\lim_{\lambda\to 0 }   \omega_2(\Phi(f_\lambda)\Phi(f'_\lambda)) =
	\lim_{\epsilon\to 0^+} - \frac{\kappa^2}{128\pi}   \int_{\mathcal{R}^4\times\mathcal{B}} d \tau dU d \tau^\prime dU^\prime dx_3\, \frac{F( \tau,U,x_3) F^\prime(\tau^\prime,U^\prime,x_3)}{(\sinh(\frac{\kappa}{2} ( \tau- \tau^\prime))+i\epsilon)^2} \ ,
	\end{equation*}
where, in view of the compactness of both $F$ and $F^\prime$ we can extend the domain of integration for the variables $\tau,\tau^\prime, U,U^\prime$ to the whole real axis, while $\mathcal{B}$ indicates a one-dimensional, connected, domain of integration, diffeomorphic to the bifurcation horizon. We can rewrite this last expression in Fourier space with respect to the variables $\tau$ and $\tau^\prime$ as
	\begin{equation*}
	\lim_{\lambda\to 0 }  \omega_2(\Phi(f_\lambda)\Phi(f^\prime_\lambda)) = \frac{1}{64} \int_{\mathbb{R}^2\times \mathcal{B}}dU dU^\prime
	d x_3 \left(\int_{-\infty}^\infty dE E\,
	\frac{\overline{\widehat{F}(E,U,x^3)}  \widehat{F}^\prime(E,U^\prime,x_3) }{1-e^{-\beta_H E}}  \right)  \;,
	\end{equation*}	
where $\beta_H=\frac{2\pi}{\kappa}$ while both $\widehat{F}$ and $\widehat{F}^\prime$ indicate the Fourier transform of $F$ and $F^\prime$ respectively. This result can be interpreted as follows: Whenever a state for a real, massive Klein Gordon field, is such that its two-point function is of Hadamard form in a geodesic neighbourhood of a point of a Killing horizon, then the modes of the two-point correlation function, built with respect to \eqref{V_to_tau}, follow in the region external to the horizon a thermal distribution at the {\em Hawking temperature} $\beta^{-1}_H$.

\vskip .2cm

\noindent{\em Case $2)$} Let us consider now two $1$-parameter families of test functions $f_\lambda$ and $f^\prime_\lambda$ obeying \eqref{eq:test_functions}, up to the replacement of $\widetilde{\mathcal{O}}$ with $\mathcal{O}_t$ for $f^\prime_\lambda$ and with $\mathcal{O}_s$ for $f_\lambda$. Following the same calculation as in the previous case the result is left unchanged except for the hyperbolic sine being replaced by the hyperbolic cosine. Eventually, rewriting the result in Fourier space with respect to the variables $\tau$ and $\tau^\prime$, we obtain
	\begin{equation}\label{eq:correlation}
	\lim_{\lambda\to 0 } \omega_2(\Phi(f_\lambda)\Phi(f^\prime_\lambda)) = \frac{1}{32} \int_{\mathbb{R}^2\times \mathcal{B}}dU dU^\prime 
	d x_3 \left(\int_{-\infty}^\infty dE E\, 
	\frac{\overline{\widehat{F}(E,U,x_3)}  \widehat{F^\prime}(E,U^\prime,x_3) }{\sinh(\beta_H E/2)} \right)  \ ,
	\end{equation}	
where, once more, $\beta_H=\frac{2\pi}{\kappa}$ while both $\widehat{F}$ and $\widehat{F}^\prime$ indicate the Fourier transform of $F$ and $F^\prime$ respectively. Since the support of the test functions and hence of the observables is located both in the interior and in the exterior region with respect to the Killing horizon, once can interpret $|\omega_2(f_\lambda,f^\prime_\lambda)|^2$ as a tunnelling probability through the horizon. By considering wave packets which are peaked around $E_0\gg 1$, \eqref{eq:correlation} yields
$$
	\lim_{\lambda\to 0 } |\omega_2(\Phi(f_\lambda),\Phi'(f'_\lambda))|^2
	\approx E_0^2 e^{-\beta_H E_0}\ \ ,
$$
which is nothing but the original result of \cite{PW:2000}.
	
\section{Hawking radiation from a BTZ black hole}
	\label{sec:BTZ}
The analysis of the previous section rests mainly on two assumptions, the existence both of a (local) Killing horizon $\mathcal{H}_+$ and of a state for a real, massive scalar field which is of Hadamard form at least in a geodesic neighbourhood of a point at $\mathcal{H}_+$. While the first requirement indicates a constraint on the underlying geometry, pointing to considering black hole spacetimes or a generalization thereof, the second one is more subtle since it refers to the underlying quantum theory. The construction of Hadamard states is a thoroughly investigated problem and it is by now established their existence when the underlying manifold is globally hyperbolic \cite{Fulling:1981cf}. In the presence of a black hole, global hyperbolicity of a domain which crosses the whole horizon is not at all an obvious feature, being satisfied in some specific scenarios such as for example the four-dimensional Schwarzschild spacetime and its Kruskal extension. In this case one can consider in a neighbourhood of the event horizon as Hadamard states for the Klein-Gordon field both the Unruh \cite{Dappiaggi_ATMP,Dappiaggi:2017kka} and the Hartle-Hawking states \cite{Sanders:2013vza,Gerard}.

In $2+1$ dimensions, the situation is far more intricate if one is interested in black hole spacetimes. As a matter of fact, in order for the underlying geometry to possess a global Killing horizon, one is forced to considering only solutions to the Einstein's equation with negative cosmological constant, that is asymptotically AdS spacetimes. The prime example is the renown BTZ spacetime \cite{Banados:1992gq}, which is not globally hyperbolic, as one can readily infer since it possesses a timelike, conformal boundary. Hence, even in the prototypical example of a $2+1$-dimensional black hole spacetime, one cannot invoke a general result to conclude that, for a real, massive scalar field, a state which is Hadamard in a neighbourhood of the horizon exists. Therefore, one of the key assumptions at the heart of the analysis in Section \ref{sec:setting} is not a priori verified and it prompts the question whether any quantum state, meeting the wanted hypotheses, exists. In this section we prove that such a pathological scenario does not occur. 

\subsection{BTZ geometry}\label{subsec:geom}

\begin{figure}[t]
	\begin{center}
		\includegraphics[scale=1]{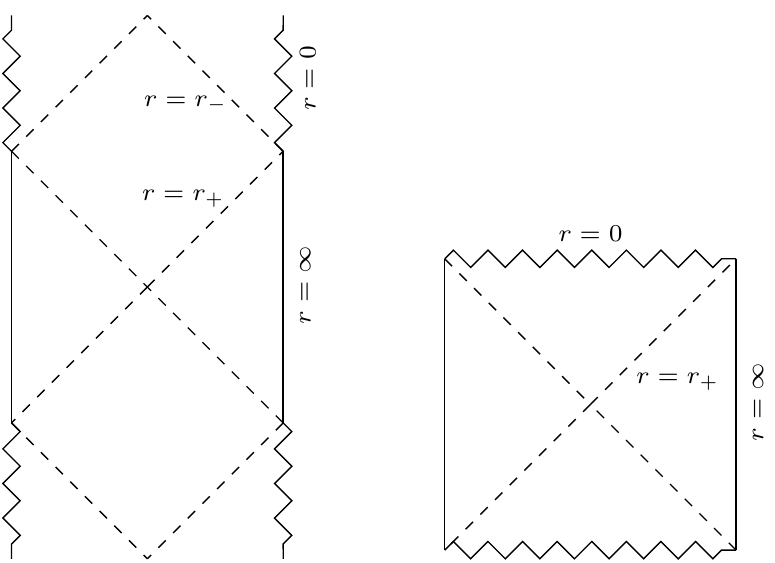}
	\end{center}
	\caption[Penrose diagrams of the BTZ.]{\label{fig:CPdiagrams} Penrose diagrams of the BTZ black hole for the rotating $0<r_-<r_+$ (left) and the static $0=r_-<r_+$ (right) cases.}
\end{figure}

In this Section, we recall the basic structural feature of the BTZ black-hole spacetime. This is a stationary, axisymmetric, (2+1)-dimensional solution of the vacuum Einstein field equations with a negative cosmological constant $\Lambda=-1/\ell^2$ \cite{Banados:1992wn,Banados:1992gq}. Henceforth we set $\ell=1$. The line element reads
\begin{equation}\label{metric}
\dd s^2 = -N^2 \dd t^2 + N^{-2} \dd r^2 + r^2 \big(\dd\phi+ N^{\phi} \dd t \big)^2\ ,
\end{equation}
where $t\in\mathbb{R}$, $\phi\in(0,2\pi)$, $r\in(0,\infty)$, while 
\begin{equation*}\label{eq:metric_functions}
N^2 = -M+r^2+\frac{J^2}{4r^2} \ , \qquad N^\phi = -\frac{J}{2r^2} \ .
\end{equation*} 
Here the parameter $M$ is interpreted as the mass of the black hole, while $J$ as its angular momentum. 
In the range  $M>0$, $|J|\le M\ell$, the values
\begin{equation}\label{eq:horizons}
r^2_{\pm}=
\frac{1}{2}
\left(
M \pm \sqrt{M^2-J^2}
\right) \ .
\end{equation}
represent an outer event horizon and an inner Cauchy horizon for the black hole. In particular the former is a a Killing horizon generated by the Killing field
\begin{equation}\label{eq:Killing_Field}
K := \partial_t +\Omega_{\mathcal{H}} \partial_\phi \ ,
\end{equation}
where $\Omega_{\mathcal{H}} := N^\phi(r_{+}) = \frac{r_{-}}{r_{+}}$ can be interpreted as the angular velocity of the horizon itself. With reference to the analysis of section \ref{sec:setting}, we will identify $\mathcal{O}_t^{BTZ}$ as per \eqref{eq:regions} as the whole region $r>r_+$, while $\mathcal{O}_s^{BTZ}$ as that for which $r_-<r<r_+$. A characteristic feature of BTZ spacetime, which distinguish it from other rotating black holes, such as for example Kerr, is that $K$ is a timelike Killing vector field across the whole $\mathcal{O}_t^{BTZ}$. A further, relevant, geometric quantity is the \textit{surface gravity} \cite[Ch. 12]{Wald:1984rg} which, in the case at hand, can be computed from the defining equation $K^\mu\nabla_\mu K_\nu=\kappa K_\nu$, K as in \eqref{eq:Killing_Field}, to be
\begin{equation*}
\kappa=\frac{r_{+}^2-r_{-}^2}{r_{+}} \ .
\end{equation*}
We stress that the BTZ black hole spacetime is not globally hyperbolic. In fact, the timelike surface $r=\infty$ behaves as a conformal boundary for $(\mathcal{M},g)$. This implies that the solutions of the equations of motion for a free field theory must be obtained by imposing suitable boundary conditions at the boundary.
The Penrose diagrams of this spacetime are shown in Fig.~\ref{fig:CPdiagrams}. 

%%%%%%%%%%%%%%%%

%%%%%%%%%%%%%%%%

As a final remark we stress that BTZ spacetime is a manifold, locally of constant, negative scalar curvature. The underlying reason is that it can be constructed via a suitable periodic identification of points, directly from AdS$_3$, the three dimensional anti-de Sitter, maximally symmetric solution to the vacuum Einstein's equations with negative cosmological constant. Since this feature will play a distinguished role in our analysis, we review the constructive procedure of BTZ spacetime succinctly, following \cite[Ch. 12]{Carlip}. 

The starting point is $\mathbb{R}^4$ endowed with Cartesian coordinates $X_i$, $i=0,...,3$ and with the line elements $ds^2=-dX^2_0-dX^2_1+dX^2_2+dX^2_3$. AdS$_3$ is realized thus as the hyperboloid 
$$-X^2_0-X^2_1+X^2_2+X^2_3=-1,$$
where the cosmological constant $\Lambda$ has been normalized to $-1$. Starting from this realization of anti-de Sitter spacetime, it is possible to cover the whole BTZ solution by means of three patches corresponding respectively to the regions {\em i)} $r\geq r_+$, {\em ii)} $r_-\leq r\leq r_+$ and {\em iii)} $r\leq r_-$.
For our later purposes only the first two are relevant, since we are interested in the behaviour of a real, massive, scalar field theory in a neighbourhood of $r=r_+$. Hence we make explicit only the first two set of coordinate transformations which read as follows:
\begin{equation}\label{patch_1}
\textrm{Region i)}\;\left\{
\begin{array}{ll}
X_0 = \sqrt{\alpha(r)}\cosh(r_+\widetilde{\phi}-r_- t), & X_1 = \sqrt{\alpha(r)-1}\sinh(r_+t-r_-\widetilde{\phi})\\
X_2 =\sqrt{\alpha(r)}\sinh(r_+\widetilde{\phi}-r_- t), & X_3 = \sqrt{\alpha(r)-1}\cosh(r_+t-r_-\widetilde{\phi})
\end{array}
\right. ,
\end{equation}
and
\begin{equation}\label{patch_2}
\textrm{Region ii)}\;\left\{
\begin{array}{ll}
X_0 = \sqrt{\alpha(r)}\cosh(r_+\widetilde{\phi}-r_- t), & X_1 = -\sqrt{\alpha(r)-1}\sinh(r_+t-r_-\widetilde{\phi})\\
X_2 =\sqrt{\alpha(r)}\sinh(r_+\widetilde{\phi}-r_- t), & X_3 = -\sqrt{\alpha(r)-1}\cosh(r_+t-r_-\widetilde{\phi})
\end{array}
\right. .
\end{equation}

\noindent Here $(t,r)$ are the same coordinates used in \eqref{metric} while $\widetilde{\phi}\in\mathbb{R}$ and $\alpha(r)=\frac{r^2-r^2_-}{r^2_+-r^2_-}$, $r_\pm$ being as in \eqref{eq:horizons}. In order to obtain the BTZ metric we need to switch from $\widetilde{\phi}$ to $\phi$ which is obtained considering the former and imposing the periodic identification $\widetilde{\phi}\equiv\widetilde{\phi}+2\pi$. As a side remark, we observe that the region {\em i)} could be also obtained via a similar, periodic identification procedure starting from the Poincar\'e patch of AdS$_3$.

\subsection{KMS state for a massive scalar field}
	\label{subsec:KMS}

We now consider a real, massive scalar field $\Phi: \mathcal{M}\to\mathbb{R}$ satisfying the Klein-Gordon equation
\begin{equation}\label{KG}
(\Box_g - m^2-\xi R)\Phi =0\ ,
\end{equation}
where $\Box_g$ is the D'Alembert wave operator built out of \eqref{metric}, $R=-6$ is the scalar curvature, while $\xi\in\mathbb{R}$ and $m^2\geq 0$ is the mass.
Since the underlying background is not globally hyperbolic, solutions of \eqref{KG} cannot be constructed only by assigning initial data on a two-dimensional, smooth, spacelike hypersurface, but one must impose also suitable boundary conditions as $r\to\infty$. This problem has been studied thoroughly in \cite{Bussola:2017} (see also \cite{Garbarz:2017wzv} for an investigation of the static case) where \eqref{KG} has been analysed in the stationary region $r>r_+$ by considering boundary conditions of Robin type. Observe that this corresponds to \eqref{patch_1} in the previous section. The ensuing space of classical solutions of the equation of motion has been used subsequently to construct, for each admissible boundary condition, the two-point function of the associated ground state, while the ensuing algebra of observables can be constructed along the same lines used for the counterpart of a real, massive scalar field in the Poincar\'e patch of a $(d+1)-$dimensional AdS spacetime, see \cite{Dappiaggi:2017wvj}.

Here we only report the final result, leaving an interested reader to \cite{Bussola:2017} for all the technical details of the derivation. Most notably, it turns out that we need to distinguish two different scenarios, which are ruled by the parameter $\mu^2\doteq m^2-6\xi$, which, according to our conventions, is dimensionless. Introducing for convenience the new coordinate $z= \frac{r^2-r_+^2}{r^2-r_-^2}\in (0,1)$,

\begin{enumerate}
	\item If $\mu^2\geq 0$, there exists only one admissible asymptotic behaviour at the conformal boundary of the solutions of \eqref{KG}. Hence, in this case, one does not need to impose boundary conditions as $r\to\infty$ (equivalently $z\to 1$) and the two-point function of the underlying ground state reads
	\begin{align} 
	\omega_2(x,x^\prime) &= \lim_{\epsilon \to 0^+} \sum\limits_{k\in\mathbb{Z}} e^{ik\left(\widetilde{\phi}-\widetilde{\phi}^\prime\right)}\int_0^\infty \frac{\dd\widetilde{\omega}}{(2\pi)^2} \, e^{-i\widetilde{\omega}\left(\widetilde{t}-\widetilde{t}^\prime-i\epsilon\right)} \left(\frac{A}{B}-\frac{\overline{A}}{\overline{B}}\right) C \, \Psi_1(z)\Psi_1(z^\prime) \, , \label{eq:tpf1}
	\end{align}
	where $\tilde{\phi}=\phi-\Omega_{\mathcal{H}} t$, $\tilde{t}=t$, while we recall that $\Omega_{\mathcal{H}}$ is the angular velocity of the horizon. The remaining unknowns, namely $A$, $B$ and $C$ are constants while $\Psi_1$ is a function depending only on $z$ constructed out of \eqref{eq:KG} via a mode expansion. For convenience we list their form explicitly in the appendix.  	
	Observe that $\tilde{\omega}=\omega-k\Omega_{\mathcal{H}}$ represents the Fourier parameter associated to the Killing field $K$ as in \eqref{eq:Killing_Field}. Since $K$ is stationary across the whole region $r>r_+$. Therefore, having constructed \eqref{eq:tpf1} only out of positive values of $\widetilde{\omega}$, we are justified in calling $\omega_2$ the two-point correlation function of a ground state.
	
	\item If $-1<\mu^2<0$, there exists a one-parameter family of admissible boundary conditions of Robin type which can be assigned at $z=1$ and which admits an associated ground state. Calling the underlying parameter $\zeta$, it turns out that, there exists a value $\zeta_*\in (0,\frac{\pi}{2})$ such that, whenever $\zeta\in [0,\zeta_*]$, the two-point function of the underlying ground state reads
	\begin{align}
	\omega_2^\zeta(x,x^\prime) &= \lim_{\epsilon \to 0^+} \sum_{k\in\mathbb{Z}} e^{ik\left(\widetilde{\phi}-\widetilde{\phi}^\prime\right)} \int_0^{\infty} \frac{\dd\widetilde{\omega}}{(2\pi)^2} \, e^{-i\widetilde{\omega} \left(\widetilde{t}-\widetilde{t}^\prime-i\epsilon\right)}\frac{\left(A\overline{B}-\overline{A}B\right) C}{|{\cos(\zeta) B-\sin(\zeta)A}|^2} \Psi_{\zeta}(z)\Psi_{\zeta}(z') \, . \label{eq:tpf2}
	\end{align}
	where $\Psi_{\zeta}(z) = \cos(\zeta)\Psi_1(z)+\sin(\zeta)\Psi_2(z)$, where $\Psi_1$ and $\Psi_2$ are listed in the appendix. 
\end{enumerate}

It is noteworthy that, since \eqref{eq:tpf1} and \eqref{eq:tpf2} identify ground states, they are all of Hadamard form as it is proven in full generality in \cite{Sahlmann:2000fh}. As a next step, we show how to switch from \eqref{eq:tpf1} and \eqref{eq:tpf2} to an associated thermal equilibrium state, satisfying the KMS condition. To this end we apply a standard procedure, see for example \cite{HHW}. The main ingredients are \eqref{eq:tpf1} and \eqref{eq:tpf2}, the two-point correlation functions of a ground state, and the existence of a timelike Killing field $K$ as in \eqref{eq:Killing_Field} with an associated action $\alpha_{\widetilde{t}}:C^\infty_0(\mathcal{M})\to C^\infty_0(\mathcal{M})$ such that, for all $f\in C^\infty_0(\mathcal{M})$ and for all $\widetilde{t}\in\mathbb{R}$, 
$$\alpha_{\widetilde{t}}f(x)=f(\widetilde{\alpha}_{-\widetilde{t}}(x)),$$
where $\widetilde{\alpha}_{-\widetilde{t}}(x)$ indicates the flow of $x\in\mathcal{M}$ built out of the integral curves of $K$. Starting from this premise, we say that a two-point correlation function $\omega_{2,\beta}\in\mathcal{D}^\prime(\mathcal{M}\times\mathcal{M})$ satisfies the KMS condition at the inverse temperature $\beta>0$ with respect to $\alpha_{\widetilde{t}}$ if, for every $f,f^\prime\in C^\infty_0(\mathcal{M})$, it holds
\begin{equation}\label{eq:KMS}
\int\limits_{\mathbb{R}}d\widetilde{t}\,\omega_{2,\beta}(f,\alpha_{\widetilde{t}}(f^\prime))e^{-i\widetilde{\omega}\widetilde{t}}=\int\limits_{\mathbb{R}}d\widetilde{t}\,\omega_{2,\beta}(\alpha_{\widetilde{t}}(f),f^\prime)e^{-i\widetilde{\omega}(\widetilde{t}+i\beta)}.
\end{equation}

Having already established the existence of a ground state in the exterior region $\mathcal{O}_t^{BTZ}$, {\it cf.} \eqref{eq:tpf1} and \eqref{eq:tpf2}, built out of the positive frequencies $\widetilde{\omega}$, one can construct straightforwardly an associated two-point correlation function obeying \eqref{eq:KMS}. If $\mu^2\ge 0$, the integral kernel reads

\begin{align}
 \omega_{2, \beta}(x,x^\prime) &= \lim_{\epsilon \to 0^+} \sum\limits_{k\in\mathbb{Z}} e^{ik\left(\widetilde{\phi}-\widetilde{\phi}^\prime\right)}\int_0^\infty \frac{\dd\widetilde{\omega}}{(2\pi)^2} \, \left(\frac{A}{B}-\frac{\overline{A}}{\overline{B}}\right) C \,  \notag \\
 &\quad 
 \Big[
 e^{-i\widetilde{\omega}\left(\widetilde{t}-\widetilde{t}^\prime-i\epsilon\right)} 
 \frac{e^{\beta\widetilde{\omega}}}{e^{\beta\widetilde{\omega}}-1}
 +
 e^{i\widetilde{\omega}\left(\widetilde{t}-\widetilde{t}^\prime+i\epsilon\right)}
 \frac{e^{-\beta\widetilde{\omega}}}{1- e^{-\beta\widetilde{\omega}}}
 \Big]  \Psi_1(z)\Psi_1(z^\prime),\label{eq:tpf1_KMS}
 \, .
 \end{align}
 
while, if $-1<\mu^2<0$ and if $\zeta\in [0,\zeta_*]$, 

  \begin{align}
\omega_{2,\beta}^\zeta(x,x^\prime) &= \lim_{\epsilon \to 0^+} \sum_{k\in\mathbb{Z}} e^{ik\left(\widetilde{\phi}-\widetilde{\phi}^\prime\right)} \int_0^{\infty} \frac{\dd\widetilde{\omega}}{(2\pi)^2} \, \frac{\left(A\overline{B}-\overline{A}B\right) C}{|{\cos(\zeta) B-\sin(\zeta)A}|^2} \notag\\
&\quad
\Big[
e^{-i\widetilde{\omega}\left(\widetilde{t}-\widetilde{t}^\prime-i\epsilon\right)} 
\frac{e^{\beta\widetilde{\omega}}}{e^{\beta\widetilde{\omega}}-1}
+
e^{i\widetilde{\omega}\left(\widetilde{t}-\widetilde{t}^\prime+i\epsilon\right)}
\frac{e^{-\beta\widetilde{\omega}}}{1- e^{-\beta\widetilde{\omega}}}
\Big]
\Psi_{\zeta}(z)\Psi_{\zeta}(z^\prime) \, .\label{eq:tpf2_KMS}
\end{align}

We observe that, for all $\beta>0$, both \eqref{eq:tpf1_KMS} and \eqref{eq:tpf2_KMS} identify two-point correlation functions satisfying the Hadamard condition as one can infer either by using the results of \cite{Sahlmann:2000fh} or by observing that $\omega_{2, \beta}-\omega_2$ and $\omega_{2, \beta}^\zeta-\omega_2^\zeta$ are smooth functions. In order to conclude the analysis one should study if there exists a specific value of $\beta$ for which both \eqref{eq:tpf1_KMS} and \eqref{eq:tpf2_KMS} can be seen as the restriction to $r>r_+$ of the two-point correlation function of a state, which is of Hadamard form also in a neighbourhood of the outer horizon. 

In order to give an answer to this query, it suffices to outline the procedure described in \cite[Ch. 12]{Carlip}. The starting point consists of recalling that  $BTZ$ is realized form CAdS$_3$, the universal cover of AdS$_3$ via the identifications \eqref{patch_1} and \eqref{patch_2}. As a consequence, starting from the integral kernel of any two-point function in CAdS$_3$, one can build a counterpart on $BTZ$ by means of the method of images which implements the periodic identification built in \eqref{patch_1} and in \eqref{patch_2}. In \cite{Carlip} it is shown that the ensuing two-point correlation function is periodic with respect to the time variable $\widetilde{t}$ under the shift $i\beta_H$, where $\beta_H=\frac{2\pi}{T_H}=\frac{2\pi r_+}{r^2_+-r^2_-}$ is proportional to the inverse Hawking temperature of the BTZ black hole. As a by-product the restriction of such two-point correlation function, restricted to the region $r>r_+$ enjoys the KMS property at the Hawking temperature. 

In \cite{Carlip} this procedure is made explicit starting from the two-point correlation function in CAdS$_3$ for the ground state of a massless, conformally coupled real scalar field enjoying either the Dirichlet or the Neumann boundary condition. Most notably both these states are of (local) Hadamard form and such property is pertained by the counterpart built on BTZ spacetime. 

Although the analysis is valid apparently for two rather special cases, the conclusion drawn can be applied also to all cases of our interest for the following reason: 
\begin{enumerate}
	\item In \cite{Dappiaggi:2018xvw} it has been shown that one can construct the two-point correlation function associated to the ground state of any Klein-Gordon field with an arbitrary coupling to scalar curvature enjoying a Robin boundary condition. In addition, all these two-point functions are locally of Hadamard form thanks to the analysis in \cite{Sahlmann:2000fh}. 
	\item Since anti-de Sitter space and its universal cover are maximally symmetric solutions of the Einstein's equations with a negative cosmological constant, one can show that, since the two-point function of the ground state is also maximally symmetric regardless of the chosen Robin boundary condition,  the associated integral kernel depends on the spacetime points $x,x^\prime$ only via the geodesic distance $\sigma_{AdS_3}(x,x^\prime)$ \cite{Allen:1985wd}.
	\item in \cite{Carlip} it is shown that, by applying the method of images, one obtains a two-point correlation function in $BTZ$ spacetime which is periodic in $\widetilde{t}$ under a shift of $i\beta_H$. This conclusion is drawn as a consequence of the explicit form of $\sigma_{AdS_3}(x,x^\prime)$ and thus it can be applied also to any state whose two-point correlation function depends only on the spacetime points via the geodesic distance. This is the case for the ground states built in \cite{Dappiaggi:2018xvw}
\end{enumerate}

Hence, starting from the work in \cite{Dappiaggi:2018xvw}, we obtain states in BTZ spacetime which are of Hadamard form in a neighbourhood of the outer horizon. This proves that all the ingredients needed to apply the method described in Section \ref{sec:setting} exist. In addition their restriction to the region $r>r_+$ must enjoy the KMS property at the Hawking temperature besides the Robin boundary condition. Thus they must coincide with \eqref{eq:tpf1_KMS} and with \eqref{eq:tpf2_KMS}.

\section{Conclusions}
In this paper we have achieved two main results. On the one hand, in the first part of the manuscript, we generalised the method proposed by Moretti and Pinamonti \cite{Moretti:2010qd} to the $2+1$-dimensional case, showing that two-point correlation function of any state, which is of Hadamard form in a neighbourhood of a Killing horizon, exhibits a thermal behaviour in the exterior region. Moreover the spectrum of the energy modes obeys a Bose-Einstein distribution at the Hawking temperature. In addition this method also shows that the tunnelling probability through the horizon is in agreement with the results found in \cite{Angheben:2005rm}, and with the semiclassical method proposed in \cite{PW:2000}. The main difference is that this new result adopts the point of view
proper of quantum field theory in curved spacetime and it is applicable regardless of the underlying quantum state, provided that it is of Hadamard form in a neighbourhood of a bifurcate horizon. 

On the other hand, in the second part we have shown that the work hypotheses used in the previous analysis are valid in concrete cases. Here we address this problem considering a massive, real scalar field with Robin boundary conditions on a BTZ spacetime --
a $2+1$-dimensional black hole solution of Einstein's field equations with a negative cosmological
constant. This part takes advantage of the recent results in \cite{Dappiaggi:2018xvw,Bussola:2017} combined with the classic ones in \cite{Carlip}.

It is important to stress that our work seems to indicate that the analysis in \cite{Moretti:2010qd} might only be sensible to the Hadamard condition and not to the specific form of the integral kernel of the two-point function of a Hadamard state in four dimensions. It would be interesting to investigate in the future whether the findings of Section \ref{sec:tunnel} apply also for spacetime dimensions $n>4$.

% ACKNOWLEDGEMENTS

\section*{Acknowledgements}
%\begin{acknowledgments}
	We would like to thank Nicol\`o Drago for the precious and continuous discussions. We are also grateful to Igor Khavkine, Jorma Louko and Nicola Pinamonti for the useful comments.
	The work of F.~B.\ was supported by a Ph.D. fellowship of the University of Pavia, that of C.~D.\ was supported by the University of Pavia. 
%\end{acknowledgments}

% APPENDICES

\appendix

\section{Notable Quantities}
\label{apx:const}
In the following we list the constants and the functions appearing in \eqref{eq:tpf1} and in \eqref{eq:tpf2}, as calculated in \cite{Bussola:2017}.
With reference to the former, we set
\begin{align*}
	A &= \frac{\Gamma(c-1)\Gamma(c-a-b)}{\Gamma(c-a)\Gamma(c-b)} \, , \\
	\notag
	B &=\frac{\Gamma(c-1)\Gamma(a+b-c)}{\Gamma(a)\Gamma(b)} \, , \\
	\notag
	C &= \frac{1}{4(r_+^2-r_-^2)\sqrt{1+\mu^2}} \, .
\end{align*}
with
\begin{equation*}
\begin{cases}
\displaystyle a= \frac{1}{2}\left(1 + \sqrt{1 + \mu^2} - i \, \frac{\widetilde{\omega}}{r_{+}-r_{-}} + i\frac{k}{r_+}\right)\ , \\
\displaystyle b= \frac{1}{2}\left(1 + \sqrt{1 + \mu^2} - i \, \frac{\widetilde{\omega}}{r_{+} + r_{-}} - i\frac{k}{r_+}\right)\ , \\
\displaystyle c= 1 - i \, \frac{ r_+ \widetilde{\omega}}{r_{+}^2 - r_{-}^2}\ ,
\end{cases}
\end{equation*}
where $\widetilde{\omega}=\omega-k\Omega_{\mathcal{H}}$, while $\omega\in\mathbb{R}$ and $k\in\mathbb{Z}$. At the same time we define
\begin{equation*}
\left\{\begin{array}{l}
\Psi_1(z)=z^{-i\frac{r_+\widetilde{\omega}}{r^2_+-r^2_-}}(1-z)^{\frac{1}{2}(1+\sqrt{1+\mu^2})}F(a,b,c;1-z),\\
\Psi_2(z)=z^{-i\frac{r_+\widetilde{\omega}}{r^2_+-r^2_-}}(1-z)^{\frac{1}{2}(1+\sqrt{1+\mu^2})}F(a-c,c-b,c-a-b+1;1-z)
\end{array}\right. 
\end{equation*}
where $F$ is the Gaussian hypergeometric function.

% BIBLIOGRAPHY

\end{document}